\documentclass[fleqn]{llncs} \usepackage{llncs_ext}

\usepackage{amsmath,amssymb,atmath}               

\newcommand{\Half}[1][1]{\frac{#1}{2}}
\newcommand{\tHalf}[1][1]{\tfrac{#1}{2}}

\usepackage[pdftex]{graphicx}
\usepackage{verbatim,url}

\usepackage{tikz}

\begin{document}

\begin{frontmatter}

\title{Faster exon assembly by sparse spliced alignment%
}
\myself\maketitle

\begin{abstract}
Assembling a gene from candidate exons 
is an important problem in computational biology.
Among the most successful approaches to this problem
is \emph{spliced alignment}, proposed by Gelfand et al.,
which scores different candidate exon chains
within a DNA sequence of length $m$
by comparing them to a known related gene sequence
of length $n$, $m = \Theta(n)$.
Gelfand et al.\ gave an algorithm for spliced alignment 
running in time $O(n^3)$.
Kent et al.\ considered sparse spliced alignment,
where the number of candidate exons is $O(n)$,
and proposed an algorithm for this problem
running in time $O(n^{2.5})$.
We improve on this result, 
by proposing an algorithm for sparse spliced alignment
running in time $O(n^{2.25})$.
Our approach is based on a new framework of 
\emph{quasi-local string comparison}.
\end{abstract}

\end{frontmatter}

\section{Introduction}
\label{s-intro}

Assembling a gene from candidate exons 
is an important problem in computational biology.
Several alternative approaches to this problem 
have been developed over time.
Among the most successful approaches
is \emph{spliced alignment} \cite{Gelfand+:96},
which scores different candidate exon chains
within a DNA sequence
by comparing them to a known related gene sequence.
In this method, the two sequences are modelled 
respectively by strings $a$, $b$ of lengths $m$, $n$.
We usually assume that $m = \Theta(n)$.
A subset of substrings in string $a$
are marked as candidate exons.
The comparison between sequences is made by string alignment.
Gelfand et al.\ \cite{Gelfand+:96} give an algorithm for spliced alignment 
running in time $O(n^3)$.

In general, the number of candidate exons $k$ may be as high as $O(n^2)$.
The method of \emph{sparse spliced alignment} 
makes a realistic assumption that, prior to the assembly, 
the set of candidate exons undergoes some filtering,
after which only a small fraction of candidate exons remains.
Kent et al.\ \cite{Kent+:06} give an algorithm for sparse spliced alignment
that, in the special case $k=O(n)$, runs in time $O(n^{2.5})$.
For asymptotically higher values of $k$,
the algorithm provides a smooth transition in running time
to the dense case $k=O(n^2)$, 
where its running time is asymptotically equal
to the general spliced alignment algorithm of \cite{Gelfand+:96}.

In this paper, we improve on the results of \cite{Kent+:06},
by proposing an algorithm for sparse spliced alignment
that, in the special case $k=O(n)$, runs in time $O(n^{2.25})$.
Like its predecessor, the algorithm also provides 
a smooth transition in running time to the dense case.
Our approach is based on a new framework 
of \emph{quasi-local string comparison},
that unifies the semi-local string comparison from \cite{Tiskin:JDA_ACID}
and fully-local string comparison.

This paper is a sequel to paper \cite{Tiskin:JDA_ACID};
we include most of its relevant material here for completeness.
However, we omit some definitions and proofs due to space constraints,
referring the reader to \cite{Tiskin:JDA_ACID} for the details.

\section{Semi-local longest common subsequences}
\label{s-llcs}

\newcommand{\tl}{\vartriangleleft}
\newcommand{\tleq}{\trianglelefteq}

We consider strings of characters from a fixed finite alphabet,
denoting string concatenation by juxtaposition.
Given a string, we distinguish between its contiguous \emph{substrings},
and not necessarily contiguous \emph{subsequences}.
Special cases of a substring 
are \emph{a prefix} and \emph{a suffix} of a string.
Given a string $a$, we denote by $a^{(k)}$ and $a_{(k)}$ 
respectively its prefix and suffix of length $k$.
For two strings $a= \alpha_1 \alpha_2 \ldots \alpha_m$ 
and $b= \beta_1 \beta_2 \ldots \beta_n$ 
of lengths $m$, $n$ respectively,
the \emph{longest common subsequence (LCS) problem}
consists in computing the length of the longest string
that is a subsequence both of $a$ and $b$.
We will call this length the \emph{LCS score} of the strings.

We define a generalisation of the LCS problem,
which we introduced in \cite{Tiskin:JDA_ACID} 
as the \emph{all semi-local LCS problem}.
It consists in computing the LCS scores
on substrings of $a$ and $b$ as follows:
\begin{itemize}
\item the \emph{all string-substring LCS problem}:
$a$ against every substring of $b$;
\item the \emph{all prefix-suffix LCS problem}:
every prefix of $a$ against every suffix of $b$;
\item symmetrically, the \emph{all substring-string LCS problem}
and the \emph{all suffix-prefix LCS problem},
defined as above but with the roles of $a$ and $b$ exchanged.
\end{itemize}
It turns out that by considering this combination of problems 
rather than each problem separately,
the algorithms can be greatly simplified.

A traditional distinction,
especially in computational biology,
is between global (full string against full string)
and local (all substrings against all substrings)
comparison.
Our problem lies in between, hence the term ``semi-local''.
Many string comparison algorithms
output either a single optimal comparison score 
across all local comparisons,
or a number of local comparison scores 
that are ``sufficiently close'' to the globally optimal.
In contrast with this approach,
we require to output all the locally optimal comparison scores.

In addition to standard integer indices 
$\ldots, -2, -1, 0, 1, 2, \ldots$,
we use \emph{odd half-integer} indices 
$\ldots, -\Half[5], -\Half[3], -\Half[1], 
 \Half[1], \Half[3], \Half[5], \ldots$.
For two numbers $i$, $j$, we write $i \tleq j$ if $j-i \in \{0,1\}$,
and $i \tl j$ if $j-i = 1$.
We denote
\begin{gather*}
\bra{i:j} = \{i, i+1, \ldots, j-1, j\} \\
\ang{i:j} = \bigbrc{i+\tHalf[1], i+\tHalf[3], \ldots, 
                    j-\tHalf[3], j-\tHalf[1]}
\end{gather*}
To denote infinite intervals of integers and odd half-integers,
we will use $-\infty$ for $i$ and $+\infty$ for $j$ where appropriate.
For both interval types $\bra{i:j}$ and $\ang{i:j}$,
we call the difference $j-i$ interval \emph{length}.

We will make extensive use of finite and infinite matrices,
with integer elements and integer or odd half-integer indices.
A \emph{permutation matrix} is a (0,1)-matrix 
containing exactly one nonzero in every row and every column.
An \emph{identity matrix} is a permutation matrix $I$,
such that $I(i,j)=1$ if $i=j$, and $I(i,j)=0$ otherwise.
Each of these definitions applies both to finite and infinite matrices.

A finite permutation matrix
can be represented by its nonzeros' index set.
When we deal with an infinite matrix,
it will typically have a finite non-trivial \emph{core},
and will be trivial (e.g.\ equal to an infinite identity matrix)
outside of this core.
An infinite permutation matrix with finite non-trivial core
can be represented by its core nonzeros' index set.

Let $D$ be an arbitrary numerical matrix
with indices ranging over $\ang{0:n}$.
Its \emph{distribution matrix}, with indices ranging over $\bra{0:n}$,
is defined by
%
\begin{gather*}
\label{eq-distribution}
d(i_0,j_0) = \sum D(i,j) \qquad i \in \ang{i_0:n}, j \in \ang{0:j_0}
\end{gather*}
for all $i_0,j_0 \in \bra{0:n}$.
%

When matrix $d$ is a distribution matrix of $D$,
matrix $D$ is called the \emph{density matrix} of $d$.
The definitions of distribution and density matrices 
extend naturally to infinite matrices.
We will only deal with distribution matrices 
where all elements are defined and finite.

We will use the term \emph{permutation-distribution matrix}
as an abbreviation of ``distribution matrix of a permutation matrix''.

We refer the reader to \cite{Tiskin:JDA_ACID}
for the definition of \emph{alignment dag}.
In the context of the alignment dag,
a substring $\alpha_i \alpha_{i+1} \ldots \alpha_j$
\emph{corresponds} to the interval $\bra{i-1:j}$;
we will make substantial use of this correspondence in \secref{s-quasi-lcs}.

We also refer the reader to \cite{Tiskin:JDA_ACID}
for the definitions of \emph{(extended) highest-score matrix},
and of its \emph{implicit representation}.
\begin{figure}[tb]
\centering
\includegraphics[bb=155 529 358 661]{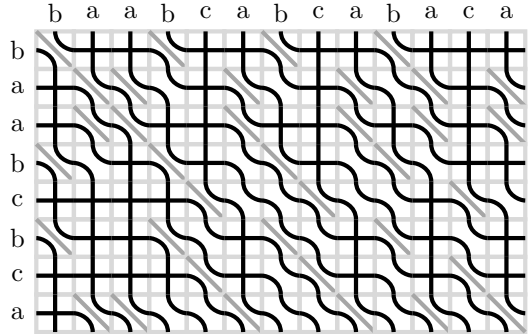}
\caption{\label{f-align-crit} 
An alignment dag and its implicit highest-score matrix}
\end{figure}
Figure~\ref{f-align-crit} shows an alignment dag of two strings,
along with the nonzeros of its implicit highest-score matrix.
In particular, a nonzero $(i,j)$, where $i,j \in \ang{0:n}$,
is represented by a ``seaweed'' curve%
\footnote{For the purposes of this illustration,
the specific layout of the curves between their endpoints is not important.
However, notice that each pair of curves have at most one crossing,
and that the same property is true 
for highest-scoring paths in the alignment dag.},
originating between the nodes $v_{0,i-\Half}$ and $v_{0,i+\Half}$,
and terminating between the nodes $v_{m,j-\Half}$ and $v_{m,j+\Half}$.
The remaining curves, originating or terminating at the sides of the dag,
correspond to nonzeros $(i,j)$,
where either $i \not\in \ang{0:n}$ or $j \not\in \ang{0:n}$.
For details, see \cite{Tiskin:JDA_ACID}.

Essentially, an extended highest-score matrix 
represents in a unified form the solutions 
of the string-substring, substring-string, 
prefix-suffix and suffix-prefix LCS problems.
In particular, row $0$ of this matrix contains the LCS scores
of string $a$ against every prefix of string $b$.
When considering such an array of $n+1$ LCS scores on its own,
we will call it \emph{highest-score vector} for $a$ against $b$.
Every highest-score vector will be represented explicitly
by an integer array of size $n+1$
(as opposed to the implicit representation 
of the complete highest-score matrix,
which allows one to store all the rows compactly 
in a data structure of size $O(m+n)$).

\section{Fast highest-score matrix multiplication}
\label{s-score-mmult}

Our algorithms are based on the framework
for the all semi-local LCS problem developed in \cite{Tiskin:JDA_ACID},
which refines the approach of \cite{Schmidt:98,Alves+:05}.

A common pattern in the problems considered in this paper
is partitioning the alignment dag into alignment subdags.
Without loss of generality, consider a partitioning 
of an $(M+m) \times n$ alignment dag $G$
into an $M \times n$ alignment dag $G_1$
and an $m \times n$ alignment dag $G_2$,
where $M \geq m$.
The dags $G_1$, $G_2$ share a horizontal row of $n$ nodes,
which is simultaneously the bottom row of $G_1$ and the top row of $G_2$;
the dags also share the corresponding $n-1$ horizontal edges.
We will say that dag $G$ is the \emph{concatenation}
of dags $G_1$ and $G_2$.
Let $A$, $B$, $C$ denote the extended highest-score matrices
defined respectively by dags $G_1$, $G_2$, $G$.
In every recursive call our goal is, given matrices $A$, $B$,
to compute matrix $C$ efficiently.
We call this procedure \emph{highest-score matrix multiplication}.

The implicit representation of matrices $A$, $B$, $C$ 
consists of respectively $M+n$, $m+n$, $M+m+n$ non-trivial nonzeros.

The results of this paper 
are based on the following results from \cite{Tiskin:JDA_ACID};
see the original paper for proofs and discussion.

\begin{mydefinition}
\label{def-minplus}
Let $n \in \Nat$. 
Let $A$, $B$, $C$ be arbitrary numerical matrices 
with indices ranging over $\bra{0:n}$.
The \emph{$(\min,+)$-product} $A \odot B = C$ is defined by
$C(i,k) = \min_j \bigpa{A(i,j) + B(j,k)}$, 
where $i,j,k \in \bra{0:n}$.
%
%
\end{mydefinition}
\begin{mylemma}[\cite{Tiskin:JDA_ACID}]
\label{lm-m-mmult}
Let $D_A$, $D_B$, $D_C$ be permutation matrices
with indices ranging over $\ang{0:n}$,
and let $d_A$, $d_B$, $d_C$ be their respective distribution matrices.
Let $d_A \odot d_B = d_C$.
Given the set of nonzero elements' index pairs in each of $D_A$, $D_B$,
the set of nonzero elements' index pairs in $D_C$
can be computed in time $O\bigpa{n^{1.5}}$ and memory $O(n)$.
\end{mylemma}
\begin{mylemma}[\cite{Tiskin:JDA_ACID}]
\label{lm-m-mmult-inf1}
Let $D_A$, $D_B$, $D_C$ be permutation matrices
with indices ranging over $\ang{-\infty:+\infty}$,
such that 
\begin{align*}
&D_A(i,j)=I(i,j) & &\text{for $i,j \in \ang{-\infty:0}$}\\
&D_B(j,k)=I(j,k) & &\text{for $j,k \in \ang{n:+\infty}$}
\end{align*}
Let $d_A$, $d_B$, $d_C$ be their respective distribution matrices.
Let $d_A \odot d_B = d_C$.
We have
\begin{align}
\label{lm-m-mmult-inf1-triv1}
&D_A(i,j)=D_C(i,j) & 
&\text{for $i \in \ang{-\infty:+\infty}$, $j \in \ang{n:+\infty}$}\\
\label{lm-m-mmult-inf1-triv2}
&D_B(j,k)=D_C(j,k) & 
&\text{for $j \in \ang{-\infty:0}$, $k \in \ang{-\infty:+\infty}$}
\end{align}
Given the set of all $n$ remaining nonzero elements' index pairs 
in each of $D_A$, $D_B$,
i.e.\ the set of all nonzero elements' index pairs 
$(i,j)$ in $D_A$ and $(j,k)$ in $D_B$
with $i \in \ang{0:+\infty}$, $j \in \ang{0:n}$, $k \in \ang{-\infty:0}$,
the set of all $n$ remaining nonzero elements' index pairs in $D_C$
can be computed in time $O\bigpa{n^{1.5}}$ and memory $O(n)$.
\end{mylemma}
\newcommand{\domain}[2][13]{%
\draw (p) ++(-0.5,0) -- (p) -- ++(#1,0) -- ++(0.5,0);
\draw (p) coordinate(pp);
\foreach \count in {0,1,...,#1}{%
  \draw (pp) node[circle](#2\count){} 
  \ifnum \count<#1 ++(1,0) coordinate(pp) \fi;}}
\newcommand{\snake}[4][]{%
\draw[#1]
  (#2) .. controls +(0,0.5) and +(0,-0.5) .. 
  (#3) .. controls +(0,0.5) and +(0,-0.5) .. (#4);}
\begin{figure}[tb]
\centering
\begin{tikzpicture}[x=0.5cm,y=-1.5cm]
\tikzstyle{every circle node}=[inner sep=0pt,minimum size=4pt,fill]
\draw (0,0) coordinate(p); \domain{I}
\draw (p) ++(0,1) coordinate(p); \domain{J}
\draw (p) ++(0,1) coordinate(p); \domain{K}
\draw (3,0) 
  +(0.5,0) node[above]{\small $0$} +(6.5,0) node[above]{\small $n$};
\draw (-0.5,0)
  +(0,0.5) node[left]{\small $D_A$} +(0,1.5) node[left]{\small $D_B$};
\snake[very thick]{I0}{J0}{K3}
\snake[very thick]{I1}{J1}{K0}
\snake[very thick]{I2}{J2}{K1}
\snake[very thick]{I3}{J3}{K4}
\snake{I4}{J6}{K2}
\snake{I5}{J8}{K9}
\snake{I6}{J4}{K6}
\snake{I7}{J5}{K8}
\snake{I8}{J9}{K7}
\snake[very thick]{I9}{J11}{K11}
\snake{I10}{J7}{K5}
\snake[very thick]{I11}{J13}{K13}
\snake[very thick]{I12}{J10}{K10}
\snake[very thick]{I13}{J12}{K12}
\end{tikzpicture}
\caption{\label{f-m-mmult-inf1} 
An illustration of \lmref{lm-m-mmult-inf1}}
\end{figure}
The lemma is illustrated by \figref{f-m-mmult-inf1}.
Three horizontal lines represent respectively 
the index ranges of $i$, $j$, $k$.
The nonzeros in $D_A$ and $D_B$ are shown respectively 
by top-to-middle and middle-to-bottom ``seaweed'' curves.
The nonzeros in $D_C$ described 
by \eqref{lm-m-mmult-inf1-triv1}, \eqref{lm-m-mmult-inf1-triv2}
are shown by top-to-bottom thick ``seaweed'' curves.
The remaining nonzeros in $D_C$ are not shown;
they are determined by application of \lmref{lm-m-mmult}
from nonzeros in $D_A$ and $D_B$
shown by top-to-middle and middle-to-bottom thin ``seaweed'' curves.

\lmref{lm-m-mmult-inf1} gives a method 
for multiplying infinite permutation-distribution matrices,
in the special case where both multiplicands have semi-infinite core.
We now consider the complementary special case,
where one multiplicand's core is unbounded,
and the other's is finite.
\begin{mylemma}
\label{lm-m-mmult-inf2}
Let $D_A$, $D_B$, $D_C$ be permutation matrices
with indices ranging over $\ang{-\infty:+\infty}$,
such that 
\begin{align*}
&D_B(j,k)=I(j,k) & &\text{for $j,k \in \ang{-\infty:0} \cup \ang{n:+\infty}$}
\end{align*}
Let $d_A$, $d_B$, $d_C$ be their respective distribution matrices.
Let $d_A \odot d_B = d_C$.
We have
\begin{align}
\label{lm-m-mmult-inf2-triv}
&D_A(i,j)=D_C(i,j) & 
&\text{for $i \in \ang{-\infty:+\infty}$, 
$j \in \ang{-\infty:0} \cup \ang{n:+\infty}$}
\end{align}
Given the set of all $n$ remaining nonzero elements' index pairs 
in each of $D_A$, $D_B$,
i.e.\ the set of all nonzero elements' index pairs 
$(i,j)$ in $D_A$ and $(j,k)$ in $D_B$
with $i \in \ang{-\infty:+\infty}$, $j,k \in \ang{0:n}$,
the set of all $n$ remaining nonzero elements' index pairs in $D_C$
can be computed in time $O\bigpa{n^{1.5}}$ and memory $O(n)$.
\end{mylemma}
\begin{myproof}
By \lmref{lm-m-mmult}; see Appendix.
\end{myproof}

\begin{figure}[tb]
\centering
\begin{tikzpicture}[x=0.5cm,y=-1.5cm]
\tikzstyle{every circle node}=[inner sep=0pt,minimum size=4pt,fill]
\draw (0,0) coordinate(p); \domain{I}
\draw (p) ++(0,1) coordinate(p); \domain{J}
\draw (p) ++(0,1) coordinate(p); \domain{K}
\draw (3,0) 
  +(0.5,0) node[above]{\small $0$} +(6.5,0) node[above]{\small $n$};
\draw (-0.5,0)
  +(0,0.5) node[left]{\small $D_A$} +(0,1.5) node[left]{\small $D_B$};
\snake[very thick]{I0}{J0}{K3}
\snake[very thick]{I1}{J1}{K0}
\snake[very thick]{I2}{J2}{K1}
\snake[very thick]{I3}{J3}{K4}
\snake{I4}{J5}{K2}
\snake{I5}{J7}{K10}
\snake{I6}{J4}{K7}
\snake{I7}{J6}{K5}
\snake{I8}{J9}{K8}
\snake{I9}{J8}{K6}
\snake[very thick]{I10}{J10}{K11}
\snake[very thick]{I11}{J11}{K9}
\snake[very thick]{I12}{J12}{K13}
\snake[very thick]{I13}{J13}{K12}
\end{tikzpicture}
\caption{\label{f-m-mmult-inf2} 
An illustration of \lmref{lm-m-mmult-inf2}}
\end{figure}
The lemma is illustrated by \figref{f-m-mmult-inf2},
using the same conventions as \figref{f-m-mmult-inf1}.


\begin{mylemma}
\label{lm-score-mmult}
Consider the concatenation of alignment dags as described above,
with highest-score matrices $A$, $B$, $C$.
Given the implicit representations of $A$, $B$,
the implicit representation of $C$
can be computed in time $O\bigpa{M+m^{0.5}n}$ and memory $O(M+n)$.
\end{mylemma}
\begin{myproof}
By \lmref{lm-m-mmult-inf2-triv}; see Appendix.
\end{myproof}

We will also need a separate efficient algorithm
for obtaining highest-score vectors
instead of full highest-score matrices.
This algorithm, which we call \emph{highest-score matrix-vector multiplication},
is complementary to the highest-score matrix multiplication algorithm
of \lmref{lm-m-mmult}.
An equivalent procedure is given
(using different terminology and notation)
in \cite{Landau+:01,Crochemore+:04,Kent+:06},
based on techniques from \cite{Kannan_Myers:96,Benson:95}.
\begin{mylemma}[\cite{Landau+:01,Crochemore+:04,Kent+:06}]
\label{lm-m-mvmult}
Let $D_A$ be a permutation matrix 
with indices ranging over $\ang{0:n}$,
and let $d_A$ be its distribution matrix.
Let $x$, $y$ be numerical (column) vectors 
with indices ranging over $\ang{0:n}$.
Let $d_A \odot x = y$.
Given the set of nonzero elements' index pairs in $D_A$,
and the elements of $x$,
the elements of $y$
can be computed in time $O\pa{n \log n}$ and memory $O(n)$.
\end{mylemma}

\section{Quasi-local string comparison}
\label{s-quasi-lcs}

Consider an arbitrary set of substrings of string $a$. 
We call substrings in this set \emph{prescribed substrings},
and denote their number by $k$.
Our aim is to compare the LCS scores 
on substrings of $a$ and $b$ as follows:
\begin{itemize}
\item the \emph{quasi-local LCS problem:}
every prescribed substring of $a$ against every substring of $b$.
\end{itemize}
This problem includes as special cases
the semi-local string comparison from \cite{Tiskin:JDA_ACID}
and fully-local string comparison,
as well as length-constrained local alignment
from \cite{Arslan_Egecioglu:02}.
Note that the solution of the quasi-local LCS problem
can be represented in space $O(kn)$
by giving the implicit highest-score matrix 
for each prescribed substring of $a$ against $b$.
An individual quasi-local LCS score query 
can be performed on this data structure in time $O(\log^2 n)$
(or even $O\bigpa{\frac{\log n}{\log\log n}}$
with a higher multiplicative constant).

In the rest of this section, we propose
an efficient algorithm for the quasi-local LCS problem.
For simplicity, we first consider the case $k=O(m)$.
Intervals corresponding to prescribed substrings of $a$
will be called \emph{prescribed intervals}.

\begin{myalgorithm}[Quasi-local LCS]
\label{alg-quasi-lcs} \setlabelitbf
\nobreakitem[Input:]
strings $a$, $b$ of length $m$, $n$, respectively;
a set of $k=O(m)$ endpoint index pairs 
for the prescribed substrings in $a$.
\item[Output:]
implicit highest-score matrix
for every prescribed substring of $a$ against full $b$.
\item[Description.]
For simplicity, we assume that $m$ is a power of $4$.
We call an interval of the form 
$\bra{k \cdot 2^s : (k+1) \cdot 2^s}$, $k,s \in \Int$, 
as well as the corresponding substring of $a$, \emph{canonical}.
In particular, all individual characters of $a$ are canonical substrings.
Every substring of $a$ can be decomposed
into a concatenation of $O(\log m)$ canonical substrings.

In the following, by processing an interval 
we mean computing the implicit highest-score matrix
for the corresponding substring of $a$ against $b$.

\setlabelit
\item[First phase.]
Canonical intervals are processed in a balanced binary tree,
in order of increasing length.
Every interval of length $2^0=1$ is canonical, 
and is processed by a simple scan of string $b$.
Every canonical interval of length $2^{s+1}$ is processed 
as a concatenation of two already processed 
half-sized canonical intervals of length $2^s$.

\item[Second phase.]
We represent each prescribed interval $\bra{i,j}$ 
by an odd half-integer \emph{prescribed point} $(i,j) \in \ang{0:m}^2$
(\footnote{%
The overall algorithm structure is essentially equivalent 
to building a one-dimensional range tree \cite{Bentley:80}
on the interval $\bra{0:m}$,
and then performing on this range tree a batch query
consisting of all the prescribed points.
However, in contrast with standard range trees,
the cost of processing nodes in our algorithm is not uniform.}).
On the set of prescribed points, we build a data structure
allowing efficient orthogonal range counting queries.
A classical example of such a data structure 
is the range tree \cite{Bentley:80}.

We then proceed by partitioning the square index pair range $\ang{0:m}^2$
recursively into regular half-sized square blocks.

Consider an $h \times h$ block
%
$\ang{i_0-h : i_0} \times \ang{j_0 : j_0+h}$.
%
The computation is organised so that 
when a recursive call is made on this block,
either we have $i_0 \geq j_0$, 
or the interval $\bra{i_0:j_0}$ is already processed.

For the current block, we query 
the number of prescribed points it contains.
If this number is zero, no further computation on the block 
or recursive partitioning is performed.
Otherwise, we have $j-i \in \{-h,0,h,2h,\ldots\}$.
If $j-i = -h$, 
then the intervals $\bra{i_0-h:j_0}$, $\bra{i_0:j_0+h}$ have length $0$, 
and the interval $\bra{i_0-h:j_0+h}$ is canonical.
If $j-i \geq 0$, we process the intervals
$\bra{i_0-h:j_0}$, $\bra{i_0:j_0+h}$, $\bra{i_0-h:j_0+h}$.
Each of these intervals can be processed by \lmref{lm-score-mmult},
appending and/or prepending a canonical interval of length $h$
to the already processed interval $\bra{i_0:j_0}$.
We then perform further partitioning of the block,
and call the procedure recursively on each of the four subblocks.

The base of the recursion is $h=1$.
At this point, we process all $1 \times 1$ blocks 
containing a prescribed point,
which is equivalent to processing the original prescribed intervals.
The computation is completed.
\setlabelitbf
\item[Cost analysis.]
\setlabelit
\item[First phase.]
The computation is dominated 
by the cost of the bottom level of the computation tree,
equal to $m/2 \cdot O(n) = O(mn)$.

\item[Second phase.]
The recursion tree has maximum degree $4$, height $\log m$, 
and $O(m)$ leaves corresponding to the prescribed points.

Consider the top-to-middle levels of the recursion tree.
In each level from the top down to the middle level,
the maximum number of nodes increases by a factor of $4$,
and the maximum amount of computation work per node
decreases by a factor of $2^{0.5}$.
Hence, the maximum amount of work per level 
increases in geometric progression,
and is dominated by the middle level $\Half[\log m]$.

Consider the middle-to-bottom levels of the recursion tree.
Since the tree has $O(m)$ leaves,
each level contains at most $O(m)$ nodes.
In each level from the middle down to the bottom level,
the maximum amount of computation work per node
still decreases by a factor of $2^{0.5}$.
Hence, the maximum amount of work per level 
decreases in geometric progression, 
and is again dominated by the middle level $\Half[\log m]$.

Thus, the computational work in the whole recursion tree
is dominated by the maximum amount of work
in the middle level $\Half[\log m]$.
This level has at most $O(m)$ nodes, each requiring at most
$O(m^{0.5} n)/2^{0.5 \cdot \Half[\log m]} = O(m^{0.25} n)$ work.
Therefore, the overall computation cost of the recursion
is at most $O(m) \cdot O(m^{0.25} n) = O(m^{1.25} n)$.
\end{myalgorithm}

The same algorithm can be applied in the case 
of general $k$, $1 \leq k \leq \binom{m}{2}$.
For $1 \leq k \leq m^{2/3}$,
the first phase dominates, so the overall computation cost is $O(mn)$.
For $m^{2/3} \leq k \leq \binom{m}{2}$,
the second phase dominates.
The dominant level in the recursion tree
will have $k$ nodes, each requiring at most
$O(m^{0.5}n/k^{0.25})$ work.
Therefore, the overall computation cost
will be at most $k \cdot O(m^{0.5}n/k^{0.25}) = O(m^{0.5} k^{0.75} n)$.
In the fully-local case $k = \binom{m}{2}$, the cost is $O(m^2 n)$;
the same result can be obtained by $m$ independent runs 
of algorithms from \cite{Schmidt:98,Alves+:05},
at the same asymptotic cost.

\section{Sparse spliced alignment}

We now consider the problem of sparse spliced alignment.
We keep the notation and terminology of the previous sections;
in particular, candidate exons are represented 
by prescribed substrings of string $a$.
We say that prescribed substring $a'=\alpha_{i'} \ldots \alpha_{j'}$
\emph{precedes} prescribed substring $a''=\alpha_{i''} \ldots \alpha_{j''}$,
if $j' < i''$.
A \emph{chain} of substrings is a chain
in the partial order of substring precedence.
We identify every chain with the string obtained
by concatenating all its constituent substrings 
in the order of precedence.

Our sparse spliced alignment algorithm is based
on the efficient method of quasi-local string comparison
developed in the previous section.
This improves the running time of the bottleneck procedure
from \cite{Kent+:06}.
The algorithm also uses a generalisation 
of the standard network alignment method,
equivalent to the one used by \cite{Kent+:06}.
For simplicity, we describe our algorithm
for the special case of unit-cost LCS score.

\begin{myalgorithm}[Sparse spliced alignment]
\label{alg-spliced} \setlabelitbf
\nobreakitem[Input:]
strings $a$, $b$ of length $m$, $n$, respectively;
a set of $k=O(m)$ endpoint index pairs 
for the prescribed substrings in $a$.
\item[Output:]
the chain of prescribed substrings in $a$,
giving the highest LCS score against string $b$.
\item[Description.]
The algorithm runs in two phases.
\setlabelit
\item[First phase.]
By running \algref{alg-quasi-lcs}, we compute 
the implicit highest-score matrix 
for every prescribed substring of $a$ against $b$.

\item[Second phase.]
We represent the problem by a \emph{dag} (directed acyclic graph)
on the set of nodes $u_i$, where $i \in\bra{0:m}$.
For each prescribed substring $\alpha_{i} \ldots \alpha_{j}$,
the dag contains the edge $u_{i-1} \to u_j$.
Overall, the dag contains $k=O(m)$ edges.

The problem can now be solved by dynamic programming 
on the representing dag as follows.
Let $s[i,j]$ denote the highest LCS score 
for a chain of prescribed substrings in prefix string $a^{(i)}$
against prefix string $b^{(j)}$.
With each node $v_i$, we associate the integer vector $s[i,\cdot]$.
The nodes are processed in increasing order of their indices.
For the node $u_0$, vector $s[0,\cdot]$ is initialised by all zeros.
For a node $u_j$, we consider every edge $u_{i-1} \to u_j$,
and compute the highest-score matrix-vector product
between vector $s[i-1,\cdot]$
and the highest-score matrix corresponding 
to prescribed string $\alpha_{i} \ldots \alpha_{j}$
by the algorithm of \lmref{lm-m-mvmult}.
Vector $s[j,\cdot]$ is now obtained
by taking the elementwise maximum between vector $s[j-1,\cdot]$ 
and all the above highest-score matrix-vector products.

The solution score is given by the value $s[m,n]$.
The solution chain of prescribed substrings can now be obtained
by tracing the dynamic programming sequence backwards
from node $u_m$ to node $u_0$. 
\setlabelitbf
\item[Cost analysis.]
\setlabelit
\item[First phase.]
\algref{alg-quasi-lcs} runs in time $O(m^{1.25} n)$.

\item[Second phase.]
For each of the $k=O(m)$ edges in the representing dag,
the algorithm of \lmref{lm-m-mvmult} runs in time $O(n \log n)$.
Therefore, the total cost of this phase 
is $O(m) \cdot O(n \log n) = O(mn \log n)$.

The overall cost of the algorithm is dominated 
by the cost of the first phase, equal to $O(m^{1.25} n)$.
\end{myalgorithm}

In the case of general $k$,
the analysis of the previous section
can be applied to obtain a smooth transition 
between the sparse and dense versions of the problem.

By a constant-factor blow-up of the alignment dag,
our algorithms can be extended from the LCS score 
to the more general edit score,
where the insertion, deletion and substitution costs
are any constant rationals.

\section{Conclusions}

We have presented an improved algorithm for sparse spliced alignment,
running in time $O(n^{2.25})$,
and providing a smooth transition in the running time to the dense case.
A natural question is whether this running time can be further improved.

Our algorithm is based on the previously developed framework
of semi-local string comparison
by implicit highest-score matrix multiplication.
The method compares strings locally
by the LCS score, or, more generally, by an edit score 
where the insertion, deletion and substitution costs
are any constant rationals.
It remains an open question whether this framework 
can be extended to arbitrary real costs,
or to sequence alignment with non-linear gap penalties.

\section{Acknowledgement}

The problem was introduced to the author by Michal Ziv-Ukelson.


\bibliographystyle{plain}
\bibliography{auto,books,path,sort,mat_bool}

\begin{thebibliography}{10}

\bibitem{Alves+:05}
C.~E.~R. Alves, E.~N. C\'aceres, and S.~W. Song.
\newblock An all-substrings common subsequence algorithm.
\newblock {\em Electronic Notes in Discrete Mathematics}, 19:133--139, 2005.

\bibitem{Arslan_Egecioglu:02}
A.~N. Arslan and {\"O}.~E{\u g}ecio{\u g}lu.
\newblock Approximation algorithms for local alignment with length constraints.
\newblock {\em International Journal of Foundations of Computer Science},
  13(5):751--767, 2002.

\bibitem{Benson:95}
G.~Benson.
\newblock A space efficient algorithm for finding the best nonoverlapping
  alignment score.
\newblock {\em Theoretical Computer Science}, 145:357--369, 1995.

\bibitem{Bentley:80}
J.~L. Bentley.
\newblock Multidimensional divide-and-conquer.
\newblock {\em Communications of the ACM}, 23(4):214--229, 1980.

\bibitem{Crochemore+:04}
M.~Crochemore, G.~M. Landau, B.~Schieber, and M.~Ziv-Ukelson.
\newblock Re-use dynamic programming for sequence alignment: An algorithmic
  toolkit.
\newblock In {\em String Algorithmics}, volume~2 of {\em Texts in
  Algorithmics}. King's College Publications, 2004.

\bibitem{Gelfand+:96}
M.~S. Gelfand, A.~A. Mironov, and P.~A. Pevzner.
\newblock Gene recognition via spliced sequence alignment.
\newblock {\em Proceedings of the National Academy of Sciences of the USA},
  93(17):9061--9066, 1996.

\bibitem{Gusfield:97}
D.~Gusfield.
\newblock {\em Algorithms on Strings, Trees, and Sequences: Computer Science
  and Computational Biology}.
\newblock Cambridge University Press, 1997.

\bibitem{Kannan_Myers:96}
S.~K. Kannan and E.~W. Myers.
\newblock An algorithm for locating non-overlapping regions of maximum
  alignment score.
\newblock {\em SIAM Journal on Computing}, 25(3):648--662, 1996.

\bibitem{Kent+:06}
C.~Kent, G.~M. Landau, and M.~Ziv-Ukelson.
\newblock On the complexity of sparse exon assembly.
\newblock {\em Journal of Computational Biology}, 13(5):1013--1027, 2006.

\bibitem{Landau+:01}
G.~M. Landau and M.~Ziv-Ukelson.
\newblock On the common substring alignment problem.
\newblock {\em Journal of Algorithms}, 41(2):338--359, 2001.

\bibitem{Schmidt:98}
J.~P. Schmidt.
\newblock All highest scoring paths in weighted grid graphs and their
  application to finding all approximate repeats in strings.
\newblock {\em SIAM Journal on Computing}, 27(4):972--992, 1998.

\bibitem{Tiskin:JDA_ACID}
A.~Tiskin.
\newblock Semi-local longest common subsequences in subquadratic time.
\newblock {\em Journal of Discrete Algorithms}.
\newblock To appear, available from
  \url{http://www.dcs.warwick.ac.uk/~tiskin/pub}.

\end{thebibliography}

\appendix\clearpage

\section{Proof of \lmref{lm-m-mmult-inf2}}

\begin{myproof}[\lmref{lm-m-mmult-inf2}]
It is straightforward to check equality \eqref{lm-m-mmult-inf2-triv},
by \eqref{eq-distribution} and \defref{def-minplus}.
Informally, each nonzero of $D_C$ 
appearing in \eqref{lm-m-mmult-inf2-triv}
is obtained as a direct combination 
of a non-trivial nonzero of $D_A$ and a trivial nonzero of $D_B$.
All remaining nonzeros of $D_A$ and $D_B$ are non-trivial,
and determine collectively the remaining nonzeros of $D_C$.
However, this time the direct one-to-one relationship 
between nonzeros of $D_C$
and pairs of nonzeros of $D_A$ and $D_B$ need not hold.

Observe that all the nonzeros of $D_A$
appearing in \eqref{lm-m-mmult-inf2-triv} with $j \in \ang{-\infty:0}$
are dominated by each of the remaining nonzeros of $D_A$.
Furthermore, none of the nonzeros of $D_A$
appearing in \eqref{lm-m-mmult-inf2-triv} with $j \in \ang{n:+\infty}$
can be dominated by any of the remaining nonzeros of $D_A$.
Hence, the nonzeros 
appearing in \eqref{lm-m-mmult-inf2-triv}
cannot affect the computation of the remaining nonzeros of $D_C$.
We can therefore simplify the problem by eliminating 
all half-integer indices $i$, $j$, $k$
that correspond to nonzero index pairs $(i,j)$ and $(j,k)$ 
appearing in \eqref{lm-m-mmult-inf2-triv},
and then renumbering the remaining indices $i$,
so that their new range becomes $\ang{0:n}$
(which is already the range of $j,k$ after the elimination).
More precisely, we define permutation matrices $D'_A$, $D'_B$, $D'_C$,
with indices ranging over $\ang{0:n}$, as follows.
Matrix $D'_A$ is obtained from $D_A$
by selecting all rows $i$ with a nonzero $D_A(i,j)$, $j \in \ang{0:n}$,
and then selecting all columns that contain a nonzero
in at least one (in fact, exactly one) of the selected rows.
Matrix $D'_B$ is obtained from $D_B$
by selecting all rows $j$ and columns $k$, where $j,k \in \ang{0:n}$.
Matrix $D'_C$ is obtained from $D_C$
by selecting all rows $i$ with a nonzero $D_C(i,k)$, $k \in \ang{0:n}$,
and then selecting all columns that contain a nonzero
in at least one (in fact, exactly one) of the selected rows.
We define $d'_A$, $d'_B$, $d'_C$ accordingly.
The index order is preserved by the above matrix transformation,
so the dominance relation is not affected.
Both the matrix transformation and its inverse 
can be performed in time and memory $O(n)$.

It is easy to check that $d'_A \odot d'_B = d'_C$,
iff $d_A \odot d_B = d_C$.
Matrices $D'_A$, $D'_B$, $D'_C$
satisfy the conditions of \lmref{lm-m-mmult}.
Therefore, given the set of nonzero index pairs of $D'_A$, $D'_B$,
the set of nonzero index pairs of $D'_C$
can be computed in time $O(n^{1.5})$ and memory $O(n)$.
\end{myproof}

\section{Proof of \lmref{lm-score-mmult}}

\begin{myproof}[\lmref{lm-score-mmult}]
By \lmref{lm-m-mmult-inf2},
all but $n$ non-trivial nonzeros of $D_C$ 
can be obtained in time and memory $O(M+n)$. 
We now show how to obtain the remaining non-trivial nonzeros
in time $O\bigpa{m^{0.5}n}$, 
instead of time $O(n^{1.5})$ given by \lmref{lm-m-mmult-inf2}.

The main idea is to decompose matrix $d_B$ 
into a $(\min,+)$-product 
of permutation-distribution matrices with small core.
The decomposition is described in terms of density matrices,
and proceeds recursively.
In each recursive step, 
we define infinite permutation matrices $D'_B$, $D''_B$,
that are obtained from the density matrix $D_B$ as follows.

Recall that non-trivial nonzeros in $D_B$
belong to the index pair range $\ang{-m:n} \times \ang{0:m+n}$.
Intuitively, the idea is 
to split the range of each index into two blocks:
\begin{gather*}
\ang{-m:n}  = \bigang{-m:\tHalf[n]} \cup \bigang{\tHalf[n]:n}\\
\ang{0:m+n} = \bigang{0:\tHalf[n]}  \cup \bigang{\tHalf[n]:m+n}
\end{gather*}
Note that the splits are not uniform,
and that among the resulting four index pair blocks in $D_B$,
the block $\bigang{\tHalf[n]:n} \times \bigang{0:\tHalf[n]}$
cannot contain any nonzeros.
We process the remaining three index pair blocks individually,
gradually introducing nonzeros in matrices $D'_B$, $D''_B$
until they become permutation matrices.
Non-trivial nonzeros in $D'_B$, $D''_B$
will belong respectively to the index ranges
$\bigang{-m:\tHalf[n]} \times \bigang{0:m+\tHalf[n]}$ and
$\bigang{\tHalf[n]:m+n} \times \bigang{m+\tHalf[n]:2m+n}$.

First, we consider all nonzeros in $D_B$
with indices $(j,k) \in \bigang{-m:\tHalf[n]} \times \bigang{0:\tHalf[n]}$.
For every such nonzero, 
we introduce a nonzero in $D'_B$ at index pair $(j,k)$.
We also consider all nonzeros in $D_B$
with indices $(j,k) \in \bigang{\tHalf[n]:n} \times \bigang{\tHalf[n]:m+n}$.
For every such nonzero, 
we introduce a nonzero in $D''_B$ at index pair $(m+j,m+k)$.

Now consider all nonzeros in $D_B$
with indices $(j,k) \in \bigang{-m:\tHalf[n]} \times \bigang{\tHalf[n]:m+n}$.
There are exactly $m$ such nonzeros.
Denote their index pairs by
$(j_0,k_0)$, $(j_1,k_1)$, \ldots, $(j_{m-1},k_{m-1})$, 
where $j_0 < j_1 < \cdots < j_{m-1}$.
For each nonzero with index pair $(j_t,k_t)$,
we introduce a nonzero in $D'_B$ at index pair $\bigpa{j_t,\tHalf[n+1]+t}$,
and a nonzero in $D''_B$ at index pair $\bigpa{\tHalf[n+1]+t,m+k_t}$.

Finally, we introduce the trivial nonzeros in $D'_B$, $D''_B$
at index pairs $(j,k)$, $k-j=m$, outside the above non-trivial ranges.
The recursive step is completed.

Let $d'_B$, $d''_B$ be the distribution matrices of $D'_B$, $D''_B$.
Let $d^*_B = d'_B \odot d''_B$,
and $d^*_C = d_A \odot d^*_B = d_A \odot d'_B \odot d''_B$,
and define $D^*_B$, $D^*_C$ accordingly.
By the construction of the decomposition of $d_B$,
matrices $D_B$ and $D^*_B$ (as well as $d_B$ and $d^*_B$)
are related by a simple shift:
for all $(i,k)$, $i,k \in\ang{-\infty,+\infty}$, we have
$D_B(j,k)=D^*_B(j,k+m)$.
Consequently, matrices $D_C$ and $D^*_C$
are related by a similar shift:
for all $(i,k)$, $i,k \in\ang{-\infty,+\infty}$, we have
$D_C(i,k)=D^*_C(i,k+m)$.

The described decomposition process continues recursively,
as long as $n \geq m$.
The problem of computing matrix $d_C$
is thus reduced, up to an index shift,
to $n/m$ instances of multiplying permutation-distribution matrices.
In every instance, one of the multiplied matrices has core of size $O(m)$.
By \lmref{lm-m-mmult-inf2},
the non-trivial part of every such multiplication can be performed 
in time $O(m^{1.5})$ and memory $O(m)$.
The trivial parts of all these multiplications
can be combined into a single scan of the nonzero sets
of $D_A$, $D_B$,
and can therefore be performed in time and memory $O(M+n)$.
Hence, the whole computation 
can be performed 
in time $O\bigpa{M + (n/m) \cdot m^{1.5}} = O(M + m^{0.5}n)$ 
and memory $O(M+n)$.
\end{myproof}
\newcommand{\showdown}[1]{\genfrac{}{}{0pt}{}{#1}{\downarrow}}
\newcommand{\showup}[1]{\genfrac{}{}{0pt}{}{\uparrow}{#1}}
\begin{figure}[tb]
\centering
\begin{tikzpicture}[x=0.5cm,y=-0.5cm]
\tikzstyle{every circle node}=[inner sep=0pt,minimum size=4pt,fill]
\draw (0,0) coordinate(p) ++(0.5,0)
  node[above]{$\showdown{-m}$}
  ++(3,0) node[above]{$\showdown{0}$} 
  ++(5,0) node[above]{$\showdown{\frac{n}{2}}$}
  ++(5,0) node[above]{$\showdown{n}$}; 
\draw (p); \domain[14]{I}
\draw[very thick] (p) ++(3.5,0) rectangle +(10,3);
\draw (p) +(1.5,1.5) node[below left]{\small $D_B$}
  ++(3,3) coordinate(p); 
\draw (p); \domain[14]{K}
\draw (p) ++(0.5,0)
  node[below]{$\showup{0}$} 
  ++(5,0) node[below]{$\showup{\frac{n}{2}}$}
  ++(5,0) node[below]{$\showup{n}$} 
  ++(3,0) node[below]{$\showup{m+n}$};
\draw 
  (I0) -- (K0) 
  (I1) -- ++(2.5,2.5) .. controls +(0.5,0) and +(0,-0.5) .. (K1)
  (I2) -- ++(1.5,1.5) .. controls +(1,0) and +(0,-1) .. (K4)
  (I3) -- ++(0.5,0.5) .. controls +(2,0) and +(0,-0.5) .. (K8)
  (I4) .. controls +(0,1) and +(0,-1) .. (K2)
  (I5) .. controls +(0,1) and +(0,-1) .. (K3)
  (I6) .. controls +(0,1) and +(-2,0) .. ++(7.5,2.5) -- (K11)
  (I7) .. controls +(0,1) and +(0,-1) .. (K5)
  (I8) .. controls +(0,1) and +(0,-1) .. (K6)
  (I9) .. controls +(0,1) and +(0,-1) .. (K7)
  (I10) .. controls +(0,1) and +(0,-1) .. (K9)
  (I11) .. controls +(0,0.5) and +(-1,0) .. ++(2.5,0.5) -- (K13)
  (I12) .. controls +(0,1) and +(0,-1) .. (K10)
  (I13) .. controls +(0,1) and +(-0.5,0) .. ++(0.5,1.5) -- (K12)
  (I14) -- (K14);
\end{tikzpicture}
\caption{\label{f-score-mmult1} 
Proof of \lmref{lm-score-mmult}: the original matrix $D_B$}
\end{figure}
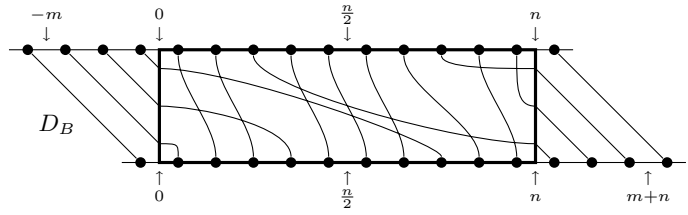
\begin{figure}[tb]
\centering
\begin{tikzpicture}[x=0.5cm,y=-0.5cm]
\tikzstyle{every circle node}=[inner sep=0pt,minimum size=4pt,fill]
\draw[fill=black!25] (8.5,0) -- ++(3,3) -- ++(0,3) -- ++(-3,-3) --cycle;
\draw (0,0) coordinate(p) ++(0.5,0)
  node[above]{$\showdown{-m}$}
  ++(3,0) node[above]{$\showdown{0}$} 
  ++(5,0) node[above]{$\showdown{\frac{n}{2}}$}
  ++(5,0) node[above]{$\showdown{n}$}; 
\draw (p); \domain[14]{I}
\draw[very thick] (p) ++(3.5,0) rectangle +(10,3);
\draw (p) +(1.5,1.5) node[below left]{\small $D'_B$}
  ++(3,3) coordinate(p); 
\draw (p); \domain[14]{J}
\draw[very thick] (p) ++(3.5,0) rectangle +(10,3);
\draw (p) +(1.5,1.5) node[below left]{\small $D''_B$}
  ++(3,3) coordinate(p); 
\draw (p); \domain[14]{K}
\draw (p) ++(0.5,0)
  node[below]{$\showup{m}$} 
  ++(5,0) node[below]{$\showup{m+\frac{n}{2}}$}
  ++(5,0) node[below]{$\showup{m+n}$} 
  ++(3,0) node[below]{$\showup{2m+n}$};
\draw 
  (I0) -- (K0) 
  (I1) -- ++(2.5,2.5) .. controls +(0.5,0) and +(0,-0.5) .. (J1) -- (K1)
  (I2) -- ++(1.5,1.5) .. controls +(1,0) and +(0,-1) .. (J4) -- (K4)
  (I3) -- ++(0.5,0.5) .. controls +(2,0) and +(-2,0) .. ++(5,2) --
    ++(3,3) .. controls +(2,0) and +(0,-0.5) .. (K8)
  (I4) .. controls +(0,1) and +(0,-1) .. (J2) -- (K2)
  (I5) .. controls +(0,1) and +(0,-1) .. (J3) -- (K3)
  (I6) .. controls +(0,1) and +(-1,0) .. ++(2.5,1.5) --
    ++(3,3) .. controls +(2,0) and +(-2,0) .. ++(5,1) -- (K11)
  (I7) .. controls +(0,1) and +(0,-1) .. (J5) -- (K5)
  (I8) .. controls +(0,0.5) and +(-0.5,0) .. ++(0.5,0.5) --
    ++(3,3) .. controls +(0.5,0) and +(0,-1) .. (K6)
  (I9) -- (J9) .. controls +(0,1) and +(0,-1) .. (K7)
  (I10) -- (J10) .. controls +(0,1) and +(0,-1) .. (K9)
  (I11) -- (J11) .. controls +(0,0.5) and +(-1,0) .. ++(2.5,0.5) -- (K13)
  (I12) -- (J12) .. controls +(0,1) and +(0,-1) .. (K10)
  (I13) -- (J13) .. controls +(0,1) and +(-0.5,0) .. ++(0.5,1.5) -- (K12)
  (I14) -- (K14);
\end{tikzpicture}
\caption{\label{f-score-mmult2} 
Proof of \lmref{lm-score-mmult}: the decomposition of $D_B$}
\end{figure}
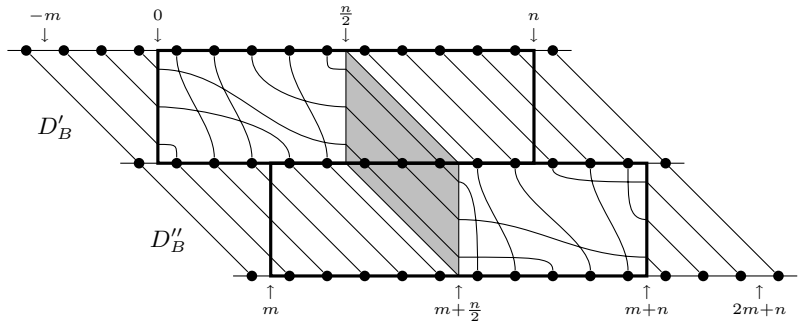
The decomposition of matrix $D_B$ in the proof of \lmref{lm-score-mmult} 
is illustrated by \figrefs{f-score-mmult1}, \ref{f-score-mmult2}.
The rectangle corresponding to $D_B$
is split into two half-sized rectangles, corresponding to $D'_B$ and $D''_B$.
Each of the new rectangles is completed 
to a full-sized rectangle by trivial extension;
then, the rectangles are arranged vertically with a shift by $m$.
The ``seaweed'' curves that do not cross the partition
are preserved by the construction, up to a shift by $m$.
The ``seaweed'' curves that cross the partition
are also preserved up to a shift by $m$,
by passing them through a parallelogram-shaped ``buffer zone''.
Note that this construction
makes the latter class of curves uncrossed in $D'_B$,
and preserves all their original crossings in $D''_B$. 

\end{document}